\documentclass[prb,twocolumn,showpacs,amsmath,amssymb]{revtex4}
\usepackage{amsfonts}

\usepackage{graphicx}
\usepackage{dcolumn}
\usepackage{bm}
\newcommand{\be}{\begin{equation}}
\newcommand{\ee}{\end{equation}}
\newcommand{\lbl}{\label}
\newcommand{\ptl}{\partial}

\def\mm#1{\mathrm{#1}}

\def\eq#1{(\ref{#1})}
\def\la{\langle}
\def\ra{\rangle}

\newcommand{\beqn}{\begin{eqnarray}}
\newcommand{\eeqn}{\end{eqnarray}}
\newcommand{\nn}{\nonumber}
\newcommand{\ai}{\emph{ab initio }}
\newcommand{\dr}{$\rho$ }

\def\eq#1{(\ref{#1})}
\def\ket#1{\vert{#1}\rangle}
\def\bra#1{\langle{#1}\vert}
\def\br#1{\langle{#1}}
\def\ord#1{{\cal O}\left({#1}\right)}
\def\hf{\frac{1}{2}}

\begin{document}


\title{Projection method for rapid ab initio calculations of metals }
\author{Abdelouaheb Kenoufi$^a$}\email{kenoufi@lpt1.u-strasbg.fr}
\author{Janos Polonyi$^{a,b}$}\email{polonyi@lpt1.u-strasbg.fr}
\affiliation{$^a$Laboratoire de Physique Th\'eorique, Universit\'e Louis
Pasteur\\
3-5 rue de l'Universit\'e F-67084 Strasbourg Cedex, France}
\affiliation{$^b$Department of Atomic Physics, Lorand E\"otv\"os University,
Budapest, Hungary}

\date{\today}

\begin{abstract}
An improvement of the Energy Renormalization Group method is proposed for
systems with small gap, based on the projection methods developed by
H. Feshbach. It is tested for the ground state energy of the
one-dimensional tight-binding model.
\end{abstract}

\maketitle

\section{Introduction}
The key point in the construction of fast, $\ord{N}$ numerical methods is
the gap in the excitation spectrum above the ground state which
renders the interactions short ranged. Systems without gap
display infrared singular, long range interactions which slow down
the convergence of the numerical algorithms. It seems natural to seek
different strategies to deal with the short and the long range quantum
fluctuations. In particular, one may use rapid numerical methods
for the short range fluctuations and treat the more difficult
long range sector with slower, more sophisticated method. Such
a mixed numerical algorithm is discussed in this paper.

The strategy of the renormalization group\cite{Wilson_1975} is a natural
candidate for the construction of such an algorithm. In fact, the
renormalization group is a systematic method to successively eliminate certain
degrees of freedom or fluctuation modes in such a manner that their impact on
the dynamics is accumulated in the effective theory which is constructed
for the remaining degrees of freedom. The algorithm proposed in this paper
consists of two steps. First, a rapid numerical method is applied for the 
elimination of the short range fluctuations. What is left is a dynamical problem
of the long range fluctuations described by an effective Hamiltonian.
This problem is dealt with in the second step of the algorithm by the 
diagonalization of the effective Hamiltonian.

The question of central importance for such a mixed method is the
relation between the total dimension of the Hilbert space and the
dimensionality of the linear space of the effective theory
where the exact diagonalization is performed. Let us denote by
$\cal E$ an intrinsic energy scale of the system and introduce
$N_>$ and $N_<$ as the number of modes with energy superior or
inferior of $\cal E$. One may call $N_>$ and $N_<$
ultraviolet and infrared cut-off. Our algorithm will be
$\ord{N_>^2}$ but will slow down as $N_<$ is
increased. Since $N_<$ growths with the size of the system
in the absence of gap and remains finite when the gap is present
our algorithm might be useful for systems with weak gap
or truly gapless models of finite size. The numerical efficiency
compared to other methods will be judged by the prefactor
of $N_>$ in the required computer time so long the system size or the
gap is kept fixed. We believe that this prefactor will be rather small
because the modes treated in this step are short ranged.

The traditional renormalization group method\cite{Wilson_1975}
consists of the repeated application of a three step procedure.
The first step is the blocking, the elimination of certain
variables from the system. This is usually achieved by
the lowering of the ultraviolet cut-off, the highest energy the
fluctuations may reach in the system. The second step is the
construction of the effective theory for the remaining modes.
Finally, in the third step which gave the name of the procedure,
one performs a rescaling of the energy or other scales of the effective
theory in order to restore the original ultraviolet cut-off.
This last step is not always necessary.

There have already been a proposal in the literature for a partial implementation
of this idea, the so called energy space renormalization
group\cite{BHG_1997_3,BHG_1998}, realizing the blocking and the
rescaling steps. In order to render this scheme systematical
one can not be content with the naive elimination of the unwanted
modes, the restriction of the Hamiltonian into a subspace, but should
realize the second step as well, the accumulation of the effects of the
excluded directions within the subspace retained. For this purpose we
use a projection method developed in nuclear physics\cite{HF_1962,Rau_1996,Muller_Rau_1996}.

In sec. \ref{dmf}, we expand the density matrix formalism, which is the
foundement of \ai algorithms.
The locality principle and its use in linear-scaling methods are presented in
sec. \ref{np}.
An exemple of such algorithm, the so called Fermi Operator Expansion is
presented in section \ref{lsa}. In the last section
we develop a Numerical Renormalization Group method in Hilbert Space around
the Fermi level and propose an improvement inspired by projection method.

\section{Density matrix formalism}\lbl{dmf}
In the framework of Density Functional Theory (DFT)
\cite{HK_1964,K_1965,Cap_2003}, particularly in the Kohn-Sham
scheme \cite{K_1965}, rapid \ai calculation methods
allowing linear scaling or near-linear scaling computation time have been
developed recently \cite{G_1999,WJ_2002}. Most of the rapid \ai algorithm
is based on the one-particle reduced density matrix \dr which is
assumed to be a projector on the subspace spanned by the
low-lying occupied states according to the \emph{auf bau} principle :
\beqn
\rho &=&
\sum\limits_{i}f_{\infty,\mu}(\epsilon_{i})\ket{\psi_{i}}\bra{\psi_{i}}\nn \\
&=&\sum\limits_{i=1}^{N/2}\ket{\psi_{i}}\bra{\psi_{i}}
\eeqn
where $N$ denotes the number of electrons and
$(\epsilon_{i},\ket{\psi_{i}})$ is an eigenfunction of the
Kohn-Sham Hamiltonian $H$ and $f_{\beta}$ is defined as the
Fermi-Dirac distribution function at the inverse temperature $\beta$,
\be
f_{\beta,\mu}(\epsilon)=\frac{1}{1+e^{\beta(\epsilon-\mu)}}.
\ee
The form $H=\sum\limits_{i}\epsilon_{i}\ket{\psi_{i}}\bra{\psi_{i}}$
allows us to write the density matrix as
\beqn\lbl{ff}
\rho &=& f_{\infty,\epsilon_{F}}(H)\nonumber\\
&=&\Theta(\epsilon_{F}-H)
\eeqn
$\Theta$ being the Heaviside function.
The average energy and the particle number can be written as
\be
E_{BS}=\sum\limits_{i,j}\rho_{ij}H_{ji}
\ee
and
\be
N=2\sum\limits_{i,j}\rho_{ij}S_{ji}
\ee
where the density matrix is given in a localized orbital basis
\be\lbl{rang}
\rho=\sum\limits_{ij}\rho_{ij}\ket{\phi_i}\bra{\phi_j}
\ee
and $H_{ij} = \bra{\phi_{i}}H\ket{\phi_{j}}$,
$S_{ij}=\br{\phi_{i}}\ket{\phi_{j}}$.

\section{Principle of "nearsightedness"}\lbl{np}
This principle states \cite{K_1959,JC_1964,K_1993} that the matrix
elements of the one-electron density matrix are negligible beyond
the distance $ca$ where $a$ is the lattice spacing,
\be\lbl{dmd2}
\vert i-j \vert>c \Rightarrow \rho_{ij}\simeq 0,
\ee
giving
\be\lbl{enp2}
E_{BS}\simeq2\sum\limits_{i}\sum\limits_{max(0,i-c)<j<min(N,i+c)}\rho_{ij}H_{ji}.
\ee

The decay of the density matrix $\rho$ in real-space depends on the
material. For systems with gap the
decay is exponential \cite{K_1959,JC_1964,K_1993,K_1995,IB}
\be
\rho(\vec{r},\vec{r}')=e^{-\alpha\vert\vec{r}-\vec{r}'\vert}
\ee
where $\alpha\sim\sqrt{\Delta\epsilon_{gap}}$ for the tight-binding limit and
$\alpha\sim a_{lattice}\cdot\Delta\epsilon_{gap}$ for the weak-binding limit.
For systems with no gap the decay of \dr is algebraic at zero
temperature\cite{IB,GI_1998,MYS}
\be
\rho(\vec{r},\vec{r}')=k_{F}\frac{\cos(k_{F}\vert\vec{r}-\vec{r}'\vert)}{\vert\vec{r}-\vec{r}'\vert^{2}}
\ee
Such an algebraic decay reflects the presence of long range correlations
and prevents linear-scaling in the numerical calculations.
The electron states tend to be more localized for disordered systems
and the matrix elements of the density matrix decay faster with
the distance\cite{Anderson,K_1996}.

\section{Fermi Operator Expansion}\lbl{lsa}
The polynomial expansion of \dr in Chebychev polynomials, the so called
Fermi Operator Expansion\cite{G_1994,G_1995,G_1998} is an important ingredient
of rapid algorithms.

Chebychev polynomials are defined by the recursion formula
\beqn\lbl{recursion}
T_{0}(x)&=&1\nn\\
T_{1}(x)&=&x \\
T_{n+2}(x)&=&2xT_{n+1}(x)-T_{n}(x)\nn
\eeqn
for $-1\le x\le1$. It is easy to see that actually
$T_{n}(x)=\cos(n\arccos x)$.
We use the functional form \eq{ff} in order to fit \dr
with a Chebychev polynomials up to order $p$,
\be
\rho_\beta=f_{\beta,\mu}(H)\simeq\sum\limits_{i=0}^pa_i(\beta_s,\mu_s)T_i(H_s)
\ee
where $H_s$ is the dimensionless Hamiltonian scaled and shifted into
the interval $[-1,1]$,
\beqn
H_{s}&=&\frac{H-\bar{E}}{\Delta E}\nn\\
\bar{E}&=&\hf[\min(spec(H))+\max(spec(H))]\\
\Delta E&=&\hf[\max(spec(H))-\min(spec(H))]\nn
\eeqn
and $\beta_{s}=\beta\Delta E$, $\mu_{s}=(\mu-\bar{E})/\Delta E$.
The smallest and largest eigenvalue of $H$ can be computed by using the
Lanczos method which scales linearly with the size of the matrix.
The projection coefficients
\be
a_n(\beta_s,\mu_s)=\frac{2-\delta_{n0}}{\pi}\int_{0}^{\pi}\cos (n
\theta)\frac{1}{1+e^{\beta_{s}(\cos \theta-\mu_{s})}}d\theta
\ee
will be computed numerically by means of Fast Fourier Transform (FFT).
The particle number conservation fixes the value of the Fermi energy level
$\epsilon_F$ found by solving Eq. \eq{rang}.

The accuracy of Fermi Operator Expansion can be estimates by recalling that
one truncates the Chebychev polynoms $T_n$ of Eq. \eq{recursion}
in such a manner that only the matrix elements $(T_{n}(H))_{ij}$ with
$\vert i-j\vert\le c$ are retained. The computation time will be of
order $pc^2N=O(N)$. It can be shown \cite{BHG_1997_1,BHG_1997_2}
that the order of the Chebychev expansion should be
\be
p\simeq\frac{2}{3}(d-1)\beta_s=\frac{2}{3}(d-1)\beta\Delta E
\ee
for the accuracy $10^{-d}$ of the expansion coefficients $\{a_{i}\}$.
If the system has an HOMO-LUMO gap $\Delta E_{gap}$ and
\be
\beta\ge\frac{2\log_{10}d}{\Delta E_{gap}}
\ee
then
\be
p\ge\frac{4(d-1)\Delta E\log_{10}d}{3\Delta E_{gap}}.
\ee
Since the range of correlations in the density matrix is bounded,
\be
range(\rho)\le p~c\simeq \frac{2}{3}c(d-1)\beta\Delta E
\ee
the correlations grows with the inverse temperature for gapless systems
and the linear-scaling methods are rendered inapplicable.

\section{Energy Renormalization Group}\lbl{erg}
We present now a renormalization group method in the energy space in order
to treat systems with small gap. In the original version of this
method\cite{BHG_1997_3,BHG_1998} one starts with a series of
inverse temperatures $\beta_n\to\infty$ and the corresponding
density matrices $\rho_n$ which tend to be concentrated around the
Fermi level as $n\to\infty$. This alone would not represent any
improvement as far as the numerical difficulties of obtaining
the density matrices are concerned. But the density matrices are
constructed in decreasing subspaces ${\cal H}_n\supset{\cal H}_{n+1}$
where ${\cal H}_{n+1}$ is span by the eigenvectors of $\rho_n$ with
large eigenvalues.

This algorithm is modified in order to implement the blocking
in energy space. First a common chemical potential is introduced
for each temperature which is adjusted at the end of the
computation to dial the desired particle number. This modification
is needed to clear the way for the blocking. The Hamiltonians
were simply truncated in the original algorithm as their subspaces
were restricted. In order to retain the dynamics of the excluded
dimensions we employ a method developed in Nuclear Physics\cite{HF_1962}
which yields an exact, $\ord{N^2}$ algorithm.

\subsection{Blocking in the Hilbert pace}
A geometric series of inverse temperatures $\beta_n=q^n\beta_0$
is introduced for $q>1$ together with the corresponding
density matrices $\rho_{n,\mu}=f_{\beta_n,\mu}(H)$.
The zero temperature expectation value of an observable $A$ is written as
a telescopic series
\be \lbl{somme}
\la A\ra=\mm{Tr(\rho_{\infty,\mu}A)}
=\sum\limits_n\mm{Tr(\Delta_{n,\mu}A)}.
\ee
where
\be
\Delta_{n,\mu}=\rho_{n,\mu}-\rho_{n-1,\mu}\nn,~~~
\Delta_{0,\mu}=\rho_{0,\mu}.
\ee
Each term in this equation corresponds to a more restricted energy subspaces
centered at the Fermi energy level as $n$ is increased. The localization
in the energy leads to delocalized states in real space in the
absence of disorder.
The ground state is approached by the telescopic summation
by zooming onto the Fermi level and the corresponding density matrix
projects on more and more extended states.

The order of the Chebychev expansion $p$ is chosen to be
independent of $n$ and the coefficients obtained by FFT are
\be
a'_m(\beta_n,\mu)=\la\Delta_{n,\mu},T_m\ra
\ee
where
\be
\Delta_{n,\mu}=f_{\beta_n,\mu}(H_n)-f_{\beta_{n-1},\mu}(H_n).
\ee

\begin{figure}[!h] \begin{center}
\includegraphics[angle=-90,width=8cm]{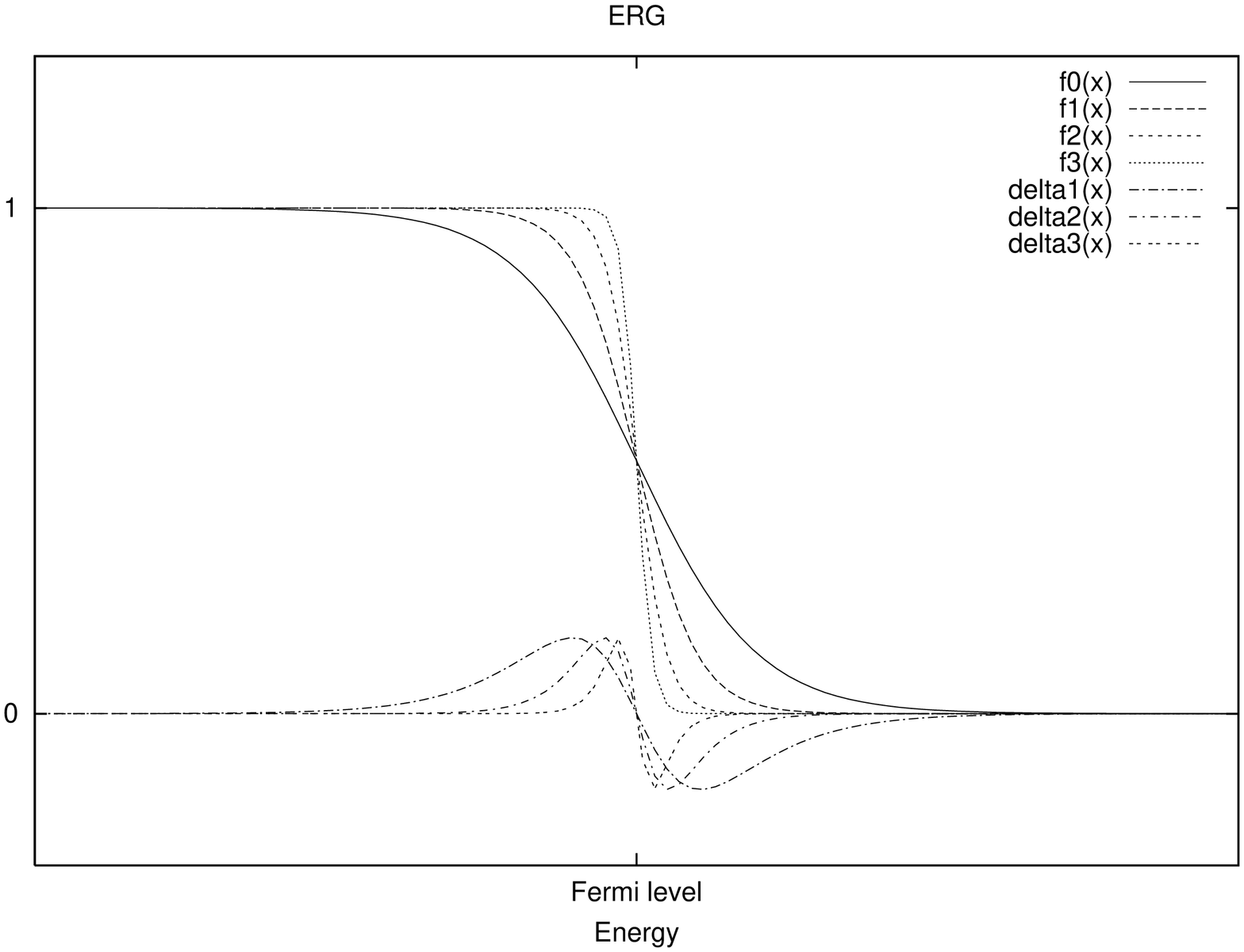}
\caption{\lbl{fig:figure1}Spectral representation of $\rho_n$ and
$\Delta_n$.}
\end{center} \end{figure}

\subsection{Fixed-point}
The convergence of the telescopic series can be expressed as the existence
of a fixed point of the blocking in the energy space for energy dependent
operators. In fact, let us suppose that a continuous operator $A$ is
commuting with the Hamiltonian $H$ and can be diagonalized in a basis of
eigenvectors of $H$. We can then express its expectation value by means
of $\rho$ as
\beqn
\mm{Tr}(\Delta_{n+1,\mu}A)&=&\int d\epsilon\ A(\epsilon)
\Delta_{n+1,\mu}(\epsilon)\nn \\
&=&\frac{\beta_n}{\beta_{n+1}}\int d\epsilon\ A
\left(\epsilon\frac{\beta_n}{\beta_{n+1}}\right)\Delta_{n,\mu}(\epsilon)\\
&=&\frac{\beta_n}{\beta_{n+1}}\mm{Tr}\left(\Delta_{n,\mu}
A\left(\cdot\frac{\beta_n}{\beta_{n+1}}\right)\right)\nonumber
\eeqn
This expression allows us to rescale the operator $A$ around the Fermi-level
by the factor $\beta_{n+1}/\beta_n$ and to keep $\Delta_{n,\mu}$ unchanged.
Since $A$ is a continuous operator the iteration of this step obviously
leads to a fixed-point,
\be
\mm{Tr}(\Delta_{n+1,\mu}A)-\mm{Tr}(\Delta_{n,\mu}A)\to0
\ee
as $n\to\infty$.

\subsection{Projection}
The identification of the subspaces proceeds by the construction of
the projectors $P_n:\mathcal{H}_n\to\mathcal{H}_{n+1}$.
We introduce first the following pseudo-projectors constructed by means
of the Chebychev expansion
\be
G_n=\frac{\ptl\rho_{n,\mu}}{\ptl\mu}=\beta_n\rho_{n,\mu}(1-\rho_{n,\mu})
\ee
Once the series $\{G_n\}$ is found another set of matrices $\{C_n\}$
is formed. The columns of $C_n$ are basis vectors of $\mathcal{H}_{n+1}$
by means of a heuristic version of the singular value decomposition with
column pivoting\cite{BHG_1997_3,BHG_1998,GVL_1996}. Since the
dynamics of the excluded dimensions is retained in our case this choice
is a less sensitive step of the algorithm then in the original
version and influences the sparsity of the resulting density matrices only.
As the next step, the overlap matrices $S_n=C_n^*C_n$ are constructed.
Finally, the projectors are given as $P_{i}=C_{i}S_{i}^{-1}C_{i}^{*}$.
$S_i^{-1}$ can actually be obtained as $S_{i}^{-1/2}=\lim_{k\to\infty}A_k$
by the help of the algorithm\cite{Lar}
\beqn\lbl{larin_1}
A_k&=&\frac{1}{2}(3A_k-A_kB_kA_k)\nn\\
B_k&=&\frac{1}{2}(3B_k-B_kA_kB_k)
\eeqn
with $A_0=-\sqrt\alpha\cdot\openone$, $B_0=-\sqrt\alpha \cdot S_{i}$
and $\alpha=1/\max\limits_{j,k}|(S_{i})_{jk}|$.
The projected Hamiltonian is of the form
\be
H_{n+1}^{ERG}=S_n^{-\hf}C_n^*H_nC_nS_n^{-\hf}
\ee

\begin{figure}[!h]
\begin{center}
\includegraphics[angle=-90,width=8cm]{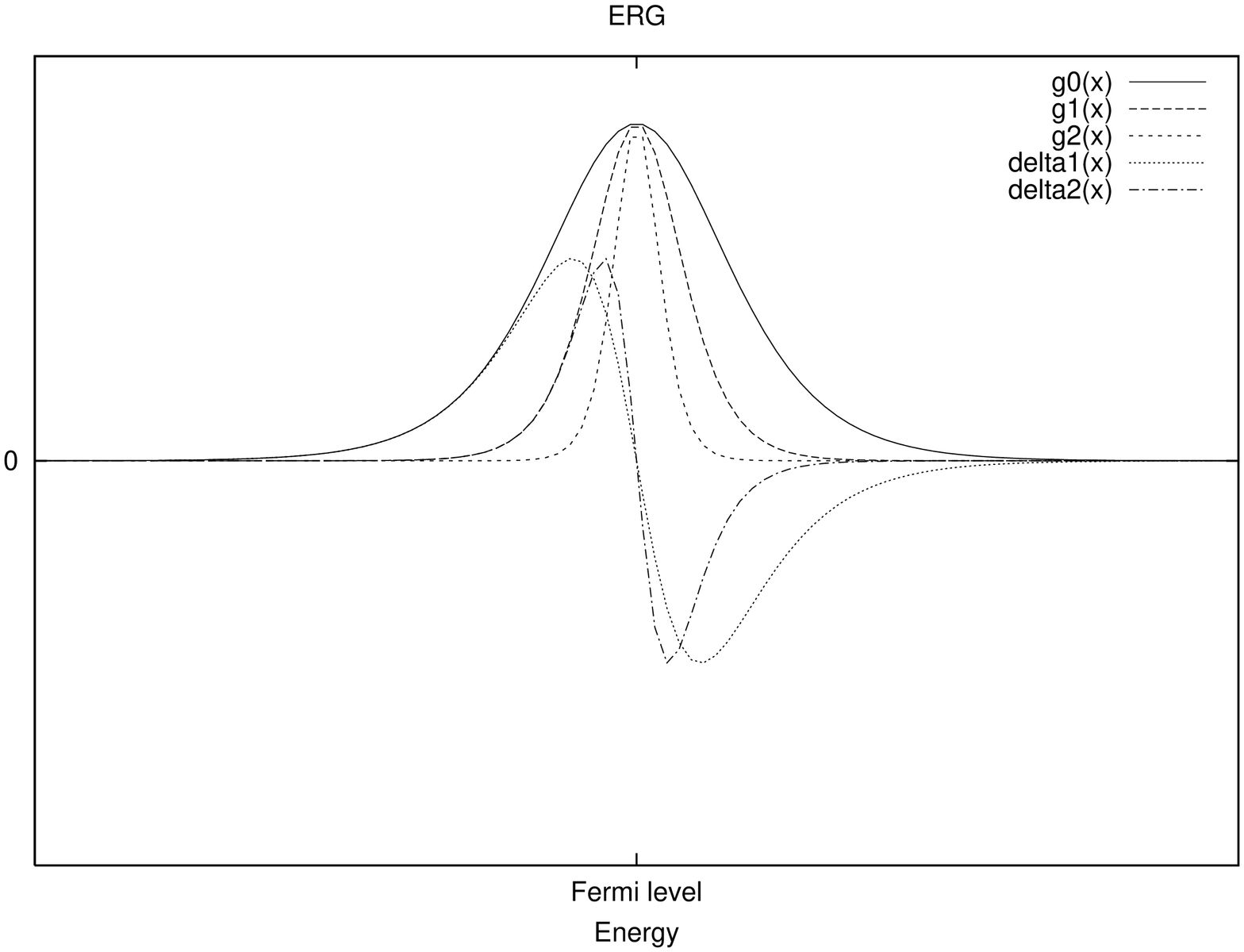}
\caption{\lbl{fig:figure2}Spectral representation of $G_{i}$ and
$\Delta_{i}$.}
\end{center}
\end{figure}

\begin{table*}
\begin{ruledtabular}
\caption{\lbl{table1} Relative errors of energy computations for different size of systems with $\beta_{0}=5$, $q=10$ and with a Chebychev
expansion $p=10$}

\begin{tabular}{cccccccc}
$N$&$E_{Exact}$&$E_{ERG}$&$E_{HF}$&$Error_{ERG}$&$Error_{HF}$&$Error_{ERG}-Error_{HF}$\\
\hline
256& 93.39 &95.44& 91.58& 2.20 &1.94& 0.26 \\
384& 139.90& 142.95 &139.09& 2.18 &0.58& 1.60 \\
512& 186.41& 190.47& 186.61& 2.18 &0.10 &2.07  \\
640 &232.93& 237.98& 234.12& 2.17 &0.51 &1.66  \\
768 &279.44& 285.50& 281.63& 2.17 &0.79 &1.38  \\
896 &325.95& 333.01& 329.15& 2.17 &0.98 &1.19  \\
1024& 372.47& 380.52& 376.66& 2.16 &1.13 &1.04  \\
1152 &418.98& 428.04& 424.17& 2.16 &1.24 &0.92  \\
1280 &465.49& 475.55& 471.69& 2.16 &1.33 &0.83  \\
1408& 512.00& 523.06& 519.20& 2.16 &1.41 &0.75  \\
1536 &558.52& 570.58& 566.72& 2.16 &1.47 &0.69  \\
1664& 605.03& 618.09& 614.23& 2.16 &1.52 &0.64  \\
1792& 651.54& 665.60& 661.74& 2.16 &1.57 &0.59  \\
1920 &698.05& 713.12& 709.26& 2.16 &1.60 &0.55  \\
2048 &744.57& 760.63& 756.77& 2.16& 1.64& 0.52 \\
\end{tabular}
\end{ruledtabular}
\end{table*}

Up to now we have excluded certain directions of the Hilbert
space which are supposed to be less important from the point
of view of the ground state dynamics. In order to perform
the analogue of the Kadanoff-Wilson blocking we have to construct
an effective Hamiltonian\cite{HF_1962,Rau_1996} $H_{n+1}$ in
the restricted space with the same dynamics around the Fermi level as
those of $H_n$,
\beqn\label{fesbach}
H_{n+1}&=&S_n^{-\hf}C_n^*\left(H_n+H_nQ_n\frac{1}{\mu-Q_nH_nQ_n}Q_nH_n\right)\nn\\
&&\times C_nS_n^{-\hf}\\
&=&H_{n+1}^{ERG}+S_n^{-\hf}C_n^*H_nQ_n\frac{1}{\mu-H_nQ_nH_n}Q_nH_nC_nS_n^{-\hf}\nn
\eeqn
where $Q_n=1-P_n$. The exclusion of directions from the Hilbert space
renders the finding of the projection of the eigenvectors of the original
Hamiltonian into the restricted space a nonlinear problem. This
complication appears as a nonlinear dependence of the eigenvector
equation in the restricted space on the eigenvalue. The energy eigenvalue
was replaced by the Fermi level, $\mu$, in the 'self-energy',
the second term on right hand side of Eq. \eq{fesbach}.
The inverse in the right hand side can be obtained
by the well-known \emph{Schultz's} or \emph{Hotelling's}
method\cite{NR,Householder,PanReif} as $(\mu-H_n)^{-1}=\lim_{j\to\infty}X_j$
where
\be\lbl{pr_1}
X_j=X_{j-1}[2\openone-(\mu-H_n)X_{j-1}]
\ee
with the initial-guess \cite{PanReif}
$X_0=(\mu-H_n)^*/\sum\limits_{j,k}(\mu-H_n)_{jk}^2$.

The calculation ends when the dimension of the
subspace is sufficiently small for explicit diagonalization.

One can introduce approximations which render the method $\ord{N}$.
One possibility is the note that $X_j$ of Eq. \eq{pr_1} converges
quadratically and the order of 30 iterations, a value independent
of the system size was always sufficient in our numerical test. Another
possibility is based on the adjustment of the chemical potential at the
end of the computation. This circumstance allows us to make the replacement
\be\label{approx}
\frac{1}{\mu-Q_nH_nQ_n}\to\frac{1}{\mu}
\ee
in Eq. \eq{fesbach} where $\mu$ will include the 'average'
of $Q_nH_nQ_n$ within ${\cal H}_n$. Such a simplification is
more acceptable for large $n$ where $\mm{dim}{\cal H}_n$ is
not too large and the evolution is slow.

\subsection{Numerical test}
We considered a lattice of $2N$ sites in one dimension with nearest
neighbor interaction described by the Hamiltonian\cite{Atkins_Friedman}
\be\label{hamnd}
H=2\sum\limits_ia_i^+a_i-\sum\limits_{<i,j>}a_i^+a_j
\ee
at half filling. Being the simplest model for the conducting band electrons the matrix elements 
of the density matrix, computed in the appendix for half-filling, show
metal-like decrease with the distance.\\ 
Table \ref{table1} shows the results of energy calculations with the
algorithm of Eq. \eq{fesbach} for different sizes.
It has been reported\cite{BHG_1997_3,BHG_1998} that the CPU time of the
ERG method scales as $N\ln^2N$. The computation of Eq. \eq{fesbach}
which was done by applying the approximation \eq{approx}
does not change this result since it contains matrix multiplications only. 

\section{Conclusion}
A new application of the renormalization group method is presented in
this work. This method is designed to retain the dynamics of modes excluded
from the computation and was developed for the path integral.
But it is an ideal tool to improve systematically the truncations
of the Hilbert space committed in the operator formalism, too.
As an example the improvement of the Energy Renormalization Group
was presented. Here the Kadanoff-Wilson blocking is performed
in energy space and the effects of the directions of the Hilbert
space lost by the truncation is retained. Therefore the dimension
of the linear space is reduced but the physics which can be described
by states within the reduced space remained the same. As long as the ground
state and the low lying excitations are kept in the linear spaces
constructed in this sequence the salient features of the model
can be described in a systematical and more economical manner.

The elimination of dimensions makes the eigenvalue equation
nonlinear in the eigenvalues, an effect which is well known in
many-body theory. In fact, say the self energy of a particle
receives a complicated, energy dependent contribution from
'virtual', particle-number changing processes which leave
from and return to the one-particle sector of the Fock space.
We employed a widely
used approximation which becomes exact for the ground state
and the low lying excitations, the replacement of the
energy eigenvalue by the Fermi level in the self-energy.
The computational need of the resulting method is $\ord{N^2_>}$
with a prefactor which growth with the volume. Nevertheless
we find this result remarkable since systems with small gap can
safely be treated by exact diagonalization in a low dimensional
subspace.

We employed a further simplification of the effective Hamiltonian
in our numerical test. We replaced the part of the Hamiltonian
which belongs to the eliminated directions and appears in the self-energy
by a 'mean-operator' which is proportional to the identity. This
approximation is supposed to become exact for the ground state and
the low lying excitations of a Fermi-liquid. The method
is $\ord{N_>\ln^2N_>}$ when this simplification is used.

Our method was tested numerically in the case of the one dimensional
tight binding model. The ground state energy improved and a
reduction of its {\em error} by 25\% was found compared to the
original algorithm for $N=2048$.

The main question left open by the present work is the dependence
of the computational requirement on $N_<$, the physical size of the system,
and the explorations of alternative approximations which ultimately
speed up the algorithm in this respect.

\begin{acknowledgments}
We thank Jean Richert, Jacques Harthong, Xavier Blanc and Claude Demangeat for useful and
interesting discussions.
\end{acknowledgments}

\appendix

\section{One-particle density matrix}
This Appendix contains some details of the computation of the density
matrix for the tight binding model of Eq. \eq{hamnd} at half filling.

The matrix of eigenvectors corresponding to the $N$ lowest eigenvalues of $H$
is given by
\be
C_{\mu\nu}=\frac{1}{\sqrt{N+\frac{1}{2}}}\sin\bigg(\frac{\pi \mu \nu}{2N+1}\bigg).
\ee
for $1\le\mu\le2N$, and $1\le\nu\le N$. The reduced density matrix
\be
\rho=CC^{*}
\ee
is a projector with the diagonal matrix elements
\be
\rho_{\mu\mu}=\hf.
\ee
If $\mu$ and $\nu$ have same parities, i.e $\mu-\nu$ ia even then
$\rho_{\mu\nu}=0$. For $\nu=\mu+2k+1$
\be
\rho_{\mu\nu}=\frac{1}{4N+2}\left[\frac{(-1)^{k}}{\sin\frac{\pi}{2}\frac{2k+1}{2N+1}}
-\frac{(-1)^{\mu+k}}{\sin\frac{\pi}{2}\frac{\mu+k+\frac{1}{2}}{2N+1}}
\right].
\ee

In order to find rate of decrease of $\rho_{\mu\nu}$ we consider the
limit $N\to\infty$ but keep $\nu-\mu=2k$ fixed,
\be
\rho_{\mu,\mu+2k}\approx\frac{(-1)^{k}}{2k\pi}.
\ee

\bibliography{apssamp}

\end{document}